\documentstyle[11pt,aasms4]{article}

\def \BD{BD +30$^\circ$ 3639}

\begin{document}

\title{Chandra X-ray Observatory Detection of 
Extended X-ray Emission from the Planetary Nebula \BD}

\author{Joel H. Kastner \\
\small
Chester F. Carlson Center for Imaging Science, Rochester Institute of 
Technology, 54 Lomb Memorial Dr., Rochester, NY 14623; jhk@cis.rit.edu \\
\normalsize
Noam Soker \\
\small
Department of Physics, University of Haifa at Oranim, Oranim, Tivon 36006, 
ISRAEL; soker@physics.technion.ac.il \\
\normalsize
Saeqa Vrtilek \\
\small
Harvard-Smithsonian Center for Astrophysics, Cambridge, MA
02138; saku@cfa.harvard.edu \\
\normalsize
Ruth Dgani \\
\small
Dept. of Astronomy, University of Texas, Austin, TX}

\begin{abstract}
We report the detection of well resolved, extended X-ray
emission from the young planetary nebula \BD\ using
the Advanced CCD Imaging Spectrometer (ACIS) aboard the Chandra
X-ray Observatory.  The X-ray emission from \BD\ appears to lie
within, but is concentrated to one side of, the interior of the shell
of ionized gas seen in high-resolution optical and IR
images. The relatively low X-ray temperature ($T_x \sim
3\times10^6$ K) and asymmetric morphology of the X-ray
emission suggests that conduction fronts are present and/or mixing of
shock-heated and photoionized gas has occurred and,
furthermore, hints at the presence of magnetic fields. The ACIS spectrum 
suggests that the X-ray emitting region is enriched in the
products of helium burning. Our detection of extended X-ray
emission from \BD\ demonstrates the power and
utility of Chandra imaging as applied to the study of
planetary nebulae. 
\end{abstract}


\section{Introduction}

Planetary nebulae (PNs) represent the last stages of evolution 
for stars of initial 
mass 1--8 $M_\odot$. The central star that generates a PN has terminated
its asymptotic giant branch (AGB)
evolution, but its ejected stellar envelope has
yet to disperse entirely, such that the star's newly exposed 
hot core (the eventual white dwarf remnant) ionizes and illuminates 
the envelope via UV radiation. The photoionized gas in a typical PN is 
far too cool to emit X-rays and the central stars are, at best,
sources of very soft X-rays.
However, theorists have long predicted that high-temperature
X-ray emission should arise from a so-called ``hot
bubble'' formed interior to the optically emitting ionized gas 
(e.g., Kwok, Purton, \& Fitzgerald, 1978; Mellema \& Frank 1995). 
The hot bubble is formed when the new, fast ($\sim1000$
km s$^{-1}$) wind emanating from the white dwarf rams into the
ejected red giant envelope that is coasting outward at $\sim10$ km s$^{-1}$,
in principle shock-heating the material to $>10^7$ K. 

{\it Einstein} and {\it ROSAT} detected X-rays
from more than a dozen planetaries (de Korte et al.\ 1985; 
Tarafdar \& Apparao 1988; Apparao \& Tarafdar
1989; Zhang, Leahy, \& Kwok
1993; Leahy, Zhang, \& Kwok 1994; 
Rauch, Koeppen, \& Werner 1994; Chu
\& Ho 1995; Hoare et al.\ 1995; Leahy et al.\ 1996; Chu,
Chang, \& Conway, 1997; Chu, Gruendl, \& Conway 1998; 
Chu, Guerrero, \& Gruendl 2000) but, in most cases, the X-ray
spectra are quite soft ($T_x \le 2\times10^5$ K) and therefore clearly
emanate from hot central stars rather than hot bubbles 
(Guerrero, Chu, \& Gruendl 2000 and references therein). But a 
handful of planetaries do display rather hard spectra,
in data obtained with the {\it ROSAT} Position-Sensitive Proportional 
Counter (PSPC). The brightest, 
hardest, and therefore most convincing example among the {\it ROSAT} PSPC
detections was the compact young planetary 
nebula BD $+30^\circ$ 3639 (Kreysing et al. 1992), whose high X-ray
temperature ($T_x \sim 3\times10^6$ K) was 
subsequently confirmed by {\it ASCA} (Arnaud,  Borkowski, \&
Harrington 1996). 
Furthermore, Leahy, Kowk, \& Yin (2000) found the X-ray
emitting region of \BD\  to be possibly extended in {\it
ROSAT} High Resolution Imager (HRI)  data.

BD $+30^\circ$ 3639
is a fascinating object in many other respects as well. 
Its youth is apparent from the compact morphology and low
ionization state of its HII region, 
the low temperature of its central star 
($T_\star \sim$ 30,000 K), and its large ratio of neutral
to ionized gas mass (Sahai \& Trauger 1998; Leahy et al.\
2000; and references therein). Taken together
these observations suggest that the onset of 
ionization in this nebula is a recent event.
Furthermore $\sim10^{-3}$
$M_\odot$ of its neutral gas is concentrated in a 
pair of diametrically opposed ``molecular bullets''
that have been expelled at $\sim50$ km s$^{-1}$ (Bachiller et al.\
2000). The presence of such highly confined
molecular outflows is strongly suggestive of the 
presence of a binary companion to the
central star (see reviews in Kastner, Soker, \& Rappaport
2000a), and detection of these ``bullets''
called into question whether the interacting winds
model was applicable to BD $+30^\circ$ 3639 (Bachiller et
al.\ 2000).

With the launch of the {\it Chandra X-ray Observatory}, it is possible 
for the first time to obtain subarcsecond images of X-ray sources and  
simultaneously to perform moderate resolution X-ray spectroscopy, with
the Advanced CCD Imaging Spectrometer (ACIS). In this {\it Letter}
we report the unambiguous detection of extended X-ray emission from 
\BD. In a second paper (Kastner, Vrtilek, \& Soker 2000b) we will
present more detailed analysis of the \BD\ data. 

\section{Observations and Data Reduction}

Chandra observed \BD, with ACIS as the focal plane
instrument, on 2000 March 21. 
The duration of the observation was 18.8 ks. The
Science Instrument Module was translated and the telescope was
pointed such that the telescope boresight was positioned near
the center of the spectroscopy CCD array (ACIS-S) and the
image of \BD\ appeared on the central back-illuminated CCD (device S3). 
The ACIS-S3 pixel size is $0.49''$. The Chandra
X-ray Center (CXC) carried out standard pipeline processing on the raw
ACIS event data, producing an aspect-corrected,
bias-subtracted, graded, energy-calibrated event list, limited to grade
02346 events (ASCA system). From this
list, we constructed a broad-band (0.3--10.0 keV) image and
a pulse height spectrum. The image
of \BD\ was smoothed by replacing raw 
pixel counts with the running mean in a $3\times3$ pixel
box. The spectrum was extracted using CXC software, which
was used to construct a histogram
of pulse heights for events contained within a circle of radius 10 pixels
(a radius we judged to include all of the X-ray flux from
\BD ). The broad-band ACIS-S3 count rate
within this aperture was 0.244 counts sec$^{-1}$. The
background count rate in a nearby, off-source region of equivalent
area was negligible in comparison ($\sim0.002$ counts sec$^{-1}$).

\section{Discussion}

The Chandra image of \BD\
shows an extended region of X-ray emission that appears to lie
within, and is concentrated to one side of, the interior of
the elliptical
shell of ionized gas seen in high-resolution optical and IR images
(Figs.\ 1, 2). The approximate extent of the X-ray
emitting region is $5''\times4''$, or $\sim5000\times4000$ AU
(for an assumed distance of $\sim1$ kpc; Kastner et al.\ 2000b).
However, the X-ray emission is quite asymmetric in comparison
to the optical and infrared morphologies of BD +30 3639,
with a compact ``hot spot'' apparent toward the
northeast rim of the optical/IR  
nebulosity and little or no emission toward the southwest
rim. This result confirms the viability of the interacting winds
model, but shows that another ingredient is necessary. In particular, the 
asymmetry suggests that the hot bubble is inhomogeneous in density, 
temperature, and/or abundance. Alternatively, a spatially varying absorbing
column may be responsible for the asymmetry; note the close 
correspondence between the outline of X-ray emission and the 
inner ``hole'' seen in the IR nebulosity.

The central star of \BD\ blows a fast wind at a velocity of
$v_f = 700$ km s$^{-1}$ (Leuenhagen, Hamann, \& Jefferyet
1996).  The shocked fast wind is expected to have a
temperature of $\sim 10^7$ K, and be distributed
isotropically around the central star. However, we find here
a much lower temperature (see below), and the X-ray emission is
localized and asymmetric, suggesting that the X-ray emission
results from cooler, clumpy gas. Such cooler and denser gas
can result from mixing of the hot gas with cooler, nebular
gas (Chu \& Ho 1995; Arnaud et al.\ 1996) and/or from the
presence of a heat conduction front (Soker 1994). The latter
possibility suggests that magnetic fields play an important
role in determining the asymmetric X-ray emission morphology
of \BD, since the presence of localized magnetic fields
would suppress heat conduction, and therefore X-ray
emission, in specific regions of the inner nebula (Soker
1994). More detailed analysis is postponed to our next paper
(Kastner et al.\ 2000b).

The Chandra/ACIS spectrum of BD $+30^\circ$ 3639 (Fig.\ 3)
shows that nearly all the X-ray photons detected by ACIS
have energies between 0.3 keV and 1.7 keV. The spectrum
displays strong emission from a blend of He-like Ne lines
centered at 915 eV, and an apparent blend of emission lines
(and, possibly, continuum emission) between $\sim300$ and
$\sim700$ eV.  Fits of a variable-abundance MEKAL
(Mewe--Kaastra--Liedahl collisional equilibrium) model
indicate that the abundance of C is enhanced relative to
solar, Ne is roughly solar, and N and O are depleted; these
fits also suggest negligible abundances for all elements
heavier than Ne.  The apparent presence of significant
continuum emission may indicate greatly enhanced He
abundance, but this result is tentative given the modest
spectral resolution of ACIS.  From the model fitting we
deduce an emitting region temperature of $T_x = 2.7\times10^6$
K ($\pm$3\%), intervening absorbing column of
$9.9\times10^{20}$ cm$^{-2}$ ($\pm$4\%), and absorbed flux
of $5.7\times10^{-13}$ ergs cm$^{-2}$ s$^{-1}$. Assuming
$D=1$ kpc, the model normalization and intrinsic
(unabsorbed) flux indicate, respectively, an emission measure
of $5.0 \times 10^{54}$ cm$^{-3}$ ($\pm$12\%) and X-ray
luminosity of $L_x \sim 1.6\times10^{32}$ ergs s$^{-1}$.  This
emission measure suggests an 
electron density $n_e \sim200$
cm$^{-3}$, given the approximate diameter of the emitting
region; as noted above, however, the asymmetry of the
Chandra image suggests the emitting region is quite clumpy,
and this result for $n_e$ should only be considered an average
value. Collectively, these results are consistent with
those obtained by Kreysing et al.\ (1992) from ROSAT data
and by Arnaud et al.\ (1996) from ASCA data, although
analysis of the latter dataset was complicated by
contamination from a nearby supernova remnant.

The ACIS spectroscopy strongly suggests that the chemical
abundances in the X-ray-emitting region are enriched in, and indeed 
may consist almost exclusively of, products of main sequence
and post-main sequence nuclear burning. 
Furthermore, the large Ne abundance suggests that the nebula
and white dwarf are the products of a relatively massive progenitor that
obtained very high core fusion temperatures and, hence, efficient
production of Ne via the reaction $^{16}$O $+$ $\alpha$ 
$\rightarrow$ $^{20}$Ne. This interpretation relies 
on the results for CNO abundances, however, which can only be tentatively 
deduced given the modest resolution of the CCD spectrum. 
The abundance anomalies implied by the X-ray spectrum may also
hint at the central star's binary nature (Soker \& Rappaport 2000).

\acknowledgements{The authors wish to acknowledge the
extraordinary efforts of the Chandra X-ray Observatory
Project and, in particular, the contributions made to the
Project by the staff of the Chandra X-ray
Center. Support for this research was provided by a
NASA/Chandra grant to RIT. N.S. also thanks the US-Israel
Binational Science Foundation. F. Roddier \&
C. Roddier kindly provided the portion of Figure 1
containing the HST and Gemini adaptive optics
images of BD$+$ 30$^\circ$3639. R. Sahai kindly
provided his fully reduced HST image of BD$+$ 30$^\circ$3639.}

\subsection*{Figure Captions}

\begin{description}
\item[Figure 1.] Optical (left), infrared (center)  
and X-ray (Chandra/ACIS; right)
images of the planetary nebula BD $+30^\circ$
3639. The optical image was obtained by HST/WFPC2 in [S III]
at 9532\AA\ (Arnaud et al.\ 1996). The infrared image was obtained
by Gemini equipped with adaptive optics camera
(http://www.ifa.hawaii.edu/ao/gemini/gemfl.html)
at 2.12 $\mu$m (H$_2$ + continuum). Images are presented at the
same spatial scale.

\item[Figure 2.] Contour map of Chandra X-ray image, overlaid on a
greyscale representation of an HST/WFPC2 6563\AA\ H$\alpha$
image (Sahai \& Trauger 1998). 
The contours correspond to total counts of 10, 30, 50, 70,
100 counts pix$^{-1}$ (integrated over the 18.8 ksec
observation).

\item[Figure 3.] Chandra/ACIS spectrum of BD $+30^\circ$ 3639
(squares), with best-fit variable-abundance MEKAL model overlaid. 
\end{description}

\end{document}